# ADRA: Extending Digital Computing-in-Memory with Asymmetric Dual-Row-Activation

Akul Malhotra, Atanu K. Saha, Chunguang Wang, and Sumeet K. Gupta
*School of Electrical and Computer Engineering, Purdue University, West Lafayette, IN 47907, USA*

*Abstract*— Computing in-memory (CiM) has emerged as an attractive technique to mitigate the von-Neumann bottleneck. Current digital CiM approaches for in-memory operands are based on multi-wordline assertion for computing bit-wise Boolean functions and arithmetic functions such as addition. However, most of these techniques, due to the many-to-one mapping of input vectors to bitline voltages, are limited to CiM of commutative functions, leaving out an important class of computations such as subtraction. In this paper, we propose a CiM approach, which solves the mapping problem through an asymmetric wordline biasing scheme, enabling (a) simultaneous single-cycle memory read and CiM of primitive Boolean functions (b) computation of any Boolean function and (c) CiM of non-commutative functions such as subtraction and comparison. While the proposed technique is technology-agnostic, we show its utility for ferroelectric transistor (FeFET)-based non-volatile memory. Compared to the standard near-memory methods (which require two full memory accesses per operation), we show that our method can achieve a full scale two-operand digital CiM using just one memory access, leading to a 23.2% - 72.6% decrease in energy-delay product (EDP).

## I. INTRODUCTION

Computing in-memory (CiM) is a promising approach for processing data-intensive workloads, for which standard von-Neumann architectures perform sub-optimally [1]. CiM (in which certain compute operations can be performed within the memory) reduces the data transfer between the processor and memory and has been extensively explored for several operations such as vector matrix multiplications, Boolean logic and arithmetic functions [2][3]. In this work, we will focus on digital CiM for Boolean and arithmetic operations.

CiM of Boolean and arithmetic functions typically utilize multi-wordline assertion to compute primitive Boolean functions (AND/OR) within the memory array. A compute unit added to the peripheral circuitry enables the computation of standard arithmetic functions such as addition. Although CiM for Boolean and arithmetic computation has been thoroughly examined, such CiM techniques are capable of computing *only commutative functions*. This is because these techniques have a *many-to-one* mapping of input vectors to senseline outputs (details discussed later). This leads to the hardware treating two different input vectors as identical, limiting the computation to only a subset of all the possibilities. For example, an important function that is not feasible using such CiM approaches is in-memory subtraction. This is because the (0,1) and (1,0) input vectors are mapped to the same senseline current/voltage. Hence, while CiM of commutative functions such as addition is possible, CiM of noncommutative functions such as subtraction and comparison in a single cycle are infeasible.

Certain other CiM techniques utilize the voltage divider action between the two words to compute their bitwise XOR and can be used for CiM of some noncommutative functions [4]. However, these techniques also suffer from the many-to-one mapping problem and are unsuitable for CiM of common functions like addition. Works like [5] demonstrate a single cycle two-bit read functionality, which could further be used to implement both commutative and noncommutative functions, but their applicability is limited to differential memories with two access transistors (e.g., SRAMs).

In this work, we propose a CiM technique which can compute *any* two-operand Boolean or arithmetic function in-memory with a single memory access. Our approach is based on asymmetric dual row activation (ADRA), which achieves *one-to-one* mapping of the input vectors to the senseline current or voltage, enabling full-scale digital CiM with in-memory operands. Although ADRA is applicable for any memory technology, we illustrate its applications employing Ferroelectric Field Effect Transistor (FeFET)-based memory. The contributions of this work are as follows:

- We propose a CiM technique, referred to as ADRA, which employs asymmetric multi-wordline assertion to avert the many-to-one mapping problem, enabling simultaneous read of two bits in a single-cycle along with CiM of any two-input Boolean function (in conjunction with a peripheral compute unit).
- We show the utility of ADRA, supported by a compute module, to perform CiM of non-commutative functions such as subtraction and comparison (in *addition* to the commutative functions explored in previous techniques).
- We evaluate ADRA considering FeFET based memory array and show that our method can achieve a full scale digital CiM with a 23.2% - 72.6% energy-delay product (EDP) decrease compared to standard CiM/near-memory methods with negligible hardware overhead.

## II. BACKGROUND

### A. Related Work

Boolean logic and arithmetic CiM using multi-wordline assertion has been presented in prior works with CMOS and emerging memories [4-7]. Fig. 1(a) illustrates the basics of this CiM technique on a generic memory array. Each bitcell produces a low resistance current ($I_{LRS}$) and a high resistance current ($I_{HRS}$) during read when it stores '1' and '0' respectively. To perform the compute operation, the bitline (RBL) and the senseline (SL) are driven to a read voltage $V_{READ}$ and ~0V, respectively. Then, wordlines (WL$_1$ and WL$_2$) corresponding to the two in-memory operands (A and B) are asserted to $V_{GREAD}$. The SL current ($I_{SL}$) is, therefore, the sum of the currents from both the bitcells. $I_{SL}$ can have three possible values for four different input vectors (see Fig. 1(c)). Using two sense amplifiers with appropriate reference currents (as shown in Fig. 1(b)), bitwise OR (A+B), bitwise AND (AB) and their complements can be computed. By adding a compute module to the peripherals (Fig. 1(d)), addition can be performed from the sense amplifier outputs.

It can be observed that $I_{SL} = (I_{HRS} + I_{LRS})$ for both the input vectors (0,1) and (1,0). Therefore, when $I_{SL}$ is equal to ($I_{HRS}$ + $I_{LRS}$), it is impossible to distinguish between (0,1) and (1,0).

This work is supported by the Center for Brain-Inspired Computing (C-BRIC), one of six centers in JUMP, funded by Semiconductor Research Corporation (SRC) and DARPA. Corresponding author: Akul Malhotra (malhot23@purdue.edu)

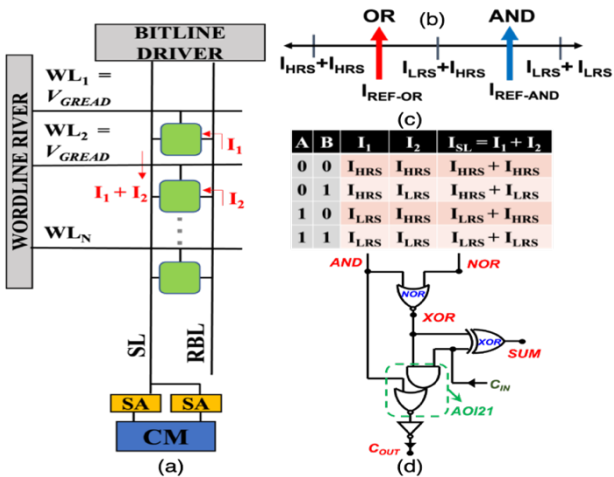

Figure 1: (a) CiM in one column of a generic memory array, (b) reference currents for the sense amplifiers, (c) $I_{SL}$ values for different input vectors and (d) Compute module for addition

This is not a problem for enabling the computation of commutative Boolean and arithmetic functions such as AND, OR, XOR and addition but prohibits the computation of *noncommutative functions* such as subtraction and comparison. The only way to distinguish between the input vectors is to perform a *second read operation* on one of them, which nearly doubles the latency and energy consumption.

An energy efficient in-memory comparison technique is proposed in [8], which needs one full and one partial memory access instead of two full memory accesses. However, it still requires two cycles and can be applied to only those non-volatile memories that can enable the partial memory access. [9] proposes a SRAM-based CAM circuit for in-memory comparison. However, the design uses a monolithic 3-D integration process which is still not mature due to manufacturing challenges such as variability. Another digital CiM technique utilizes the voltage divider action to compute the bitwise XOR of two words [4]. Although this technique can distinguish between the (0,1) and (1,0), it is unable to distinguish between the (0,0) and (1,1) input vectors, limiting the computations it can enable. Techniques like Memristor-Aided-Logic (MAGIC) enable computation in RRAM based crossbar arrays, but need many intermediate cycles to complete the computation, leading to performance and energy inefficiencies [10]. The work in [5] implements a single-cycle two-bit read by splitting the wordlines of the conventional 6T SRAM bitcell. Although this technique overcomes the previously discussed mapping problem, its use is limited to differential bitcells which have two read access transistors.

Our proposed asymmetric dual row activation (ADRA) technique overcomes the input mapping problem by mapping each of the 4 input vectors to 4 different $I_{SL}$ values. This allows us to implement any Boolean function, including subtraction and comparison in addition to those implemented in prior works. Moreover, ADRA is technology-agnostic and can be employed for all CMOS and non-volatile memories.

### B. FeFET based Non-volatile Memories (NVMs)

FeFET-based memories are amongst the most promising emerging NVMs due to their low power electric-field driven write, high distinguishability and read-write path separation [11]. These benefits come with their own challenges such as endurance issues and variability; however, many solutions are being explored to counter them [12]. In this paper, we choose FeFET NVMs to illustrate ADRA because of their aforementioned attractive features and their compatibility with current- and voltage-based sensing [13], which allows us to analyze ADRA considering both the sensing schemes.

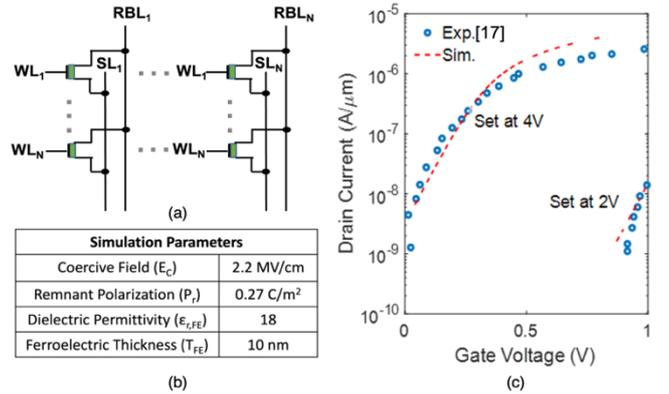

Figure 2: (a) 1T FeFET array, (b) simulation parameters and (c) I-V characteristics of the modelled device calibrated using the work in [17]

An FeFET consists of a ferroelectric (FE) layer embedded in the gate stack of a transistor (FET). Unlike dielectric materials, FEs exhibit spontaneous polarization ($P$) whose direction can be altered by applying an electric field larger than the coercive field $E_C$ directed opposite to the $P$ direction. The retention of positive and negative $P$ states ($\pm P$) in the FE layer causes a change in the threshold voltage ($V_T$) of the FeFET, which enables NVM functionalities [14].

The FeFET NVM array is composed of 1T-FeFET bitcell (Fig 2(a)). Each bitcell is connected to a senseline (SL) and a read bit line (RBL) along the column and a wordline (WL) along the row. The bit value is stored in the polarization of the FeFET. To write a value into the bitcell, an appropriate gate-to-source voltage ($V_{GS}$) is applied on the FeFET ($V_{GS} > V_C$ to write positive polarization (+P: LRS) and $V_{GS} < -V_C$ for negative polarization (-P: HRS): where $V_C$ is the coercive voltage of the FeFET). In an array, this can be accomplished in different ways such as using a two phase write or FLASH-like global reset and selective set [15]. To read the bit value, both voltage and current-based sensing can be used. For the former, the RBL is precharged to a read voltage ($V_{READ}$) with SL biased at 0V. The wordline is asserted to $V_{GREAD}$ ($<V_C$). Hence, RBL discharges for +P and remains at $V_{READ}$ for -P, which is sensed by a sense amplifier. For the current-based sensing, RBL is biased $V_{READ}$ and on the assertion of WL, the value of $I_{SL}$ ($I_{LRS}$ for +P and $I_{HRS}$ for -P) is sensed. Some previous works have explored CiM with FeFETs, albeit with the limitations described before [13].

To evaluate the FeFET arrays, we employ Preisach/Miller's equation for FE coupled with 45nm FET models [16][18]. We calibrate the FeFET model with experiments on CMOS-compatible $Hf_{0.5}Zr_{0.5}O_2$-based FeFET (Fig. 2(c)) [17]. The simulation parameters are summarized in Fig. 2(b).

### III. PROPOSED CiM USING ASYMMETRIC DUAL ROW ACTIVATION (ADRA)

#### A. Asymmetric wordline assertion scheme

As discussed in the previous section, the limitations of the previous CiM techniques are due to the many-to-one mapping of input vectors to senseline currents. In this section, we describe an asymmetric wordline assertion scheme called ADRA to alleviate this problem.

To map each input vector to a different $I_{SL}$ value, we exploit the fact that the current from each bitcell is dependent on both the bitcell value and the $V_{GREAD}$ its wordline is driven to. Therefore, if we have two different $V_{GREAD}$ values for the two wordlines, we would be able to produce unique $I_{SL}$ values for each input vector. Fig. 3(a) describes the proposed

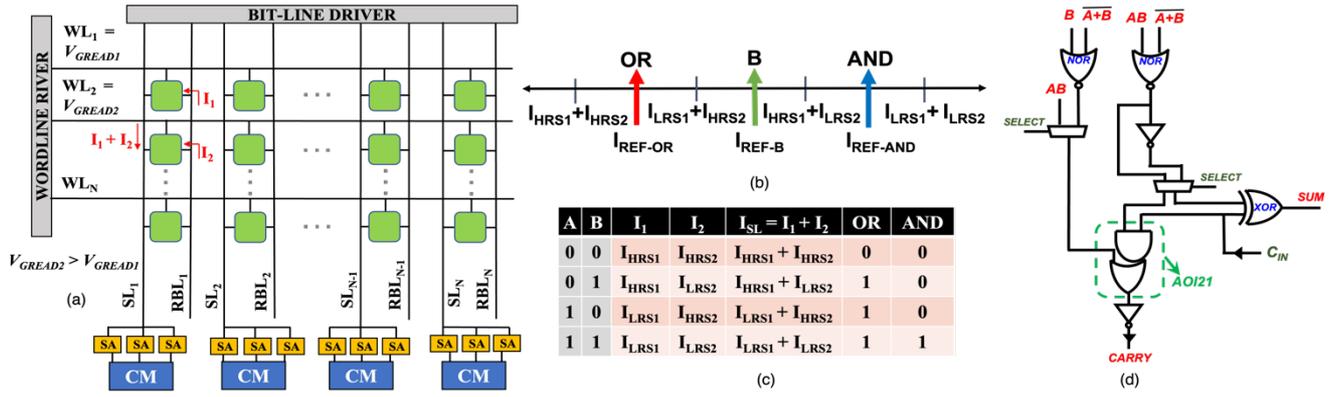

**Figure 3:** (a) ADRA based CiM in a generic memory array, (b) reference currents for the three sense amplifiers, (c) $I_{SL}$ values for different input vectors and (d) add/subtract compute module

technique using a generic memory array. During the compute operation, the RBL is driven to $V_{READ}$. The wordline of the first word (WL$_1$) is asserted to $V_{GREAD1}$, and the wordline of the second word (WL$_2$) is asserted to $V_{GREAD2}$, where $V_{GREAD2}$ > $V_{GREAD1}$. The values of $V_{GREAD1}$ and $V_{GREAD2}$ are chosen such that the difference between the $I_{SL}$ values for different input vectors is greater than the sense margin. ADRA is enabled using two row decoders (like prior CiM works [6]) with their final stages driving the voltages $V_{GREAD1}$ and $V_{GREAD2}$ respectively. Fig. 3(c) illustrates how distinct $I_{SL}$ values are obtained for each input vector. Like the prior CiM works, SL is connected to the positive input of the sense amplifiers with appropriate references (shown in Fig. 3(b)) to compute different functions. Since there are 4 different $I_{SL}$ values, an additional sense amplifier is needed (i.e., in addition to the two sense amplifiers used in standard CiM). The reference current $I_{REF-B}$ is placed between ($I_{LRS1}$ + $I_{HRS2}$) and ($I_{HRS1}$ + $I_{LRS2}$). Thus, this additional sense amplifier outputs the value of the bit 'B', which is the word in the row with the wordline voltage = $V_{GREAD2}$. The other two sense amplifiers are used to compute the bitwise OR/NOR and AND/NAND of the two words, like the prior CiM works. Likewise, ADRA can be used with voltage-based sensing by comparing the voltage discharge on the senseline to appropriate references. The additional sense amplifier required by ADRA has a minute contribution to the already large bitline capacitance, leading to negligible impact on the read/write performance.

By using the outputs of the three sense amplifiers and an additional OAI gate, A can be computed as follows:

$$A = \overline{A\overline{B}(B + \overline{A+B})}$$

Since both A and B have been sensed, a single cycle 2-bit read operation is possible with the proposed technique, along with simultaneous computation of AND and OR. Once both A and B are available, a compute module can be integrated near the memory to compute any Boolean and arithmetic functions as required. However, it is important to mention that the complexity of the compute module needs to be kept in mind while deciding the functions to be computed. Nevertheless, ADRA achieves full-scale digital CiM of functions with two in-memory-operands, and thus offers extensive flexibility in designing the CiM engine as per the needs of the system. For example, our approach enables the CiM of important class of functions, such as subtraction and comparison (details in the next sub-section), which are crucial for many mainstream and emerging workloads.

*B. In-memory subtraction and comparison*

In this section, we utilize the one-on-one mapping between input vectors and senseline currents to enable single cycle CiM of subtraction and comparison. A compact compute module which can perform both addition and subtraction is shown in Fig. 3(d). The inputs to the compute module are the outputs of the three sense amplifiers: OR (A+B), AND (AB) and B, and their compliments. Having B as an input allows for the computation of $A\overline{B}$, which is necessary for computing the subtraction (A-B). An additional 'SELECT' enables selection between the computation of addition ('SELECT' = '0') and subtraction ('SELECT' = '1'). The proposed module has additional two 2:1 multiplexers, one NOT and one NOR gate compared to the compute modules used in prior CiM works (Fig 1(d).). Our layout estimates based on scalable CMOS rules show that ADRA incurs an area increase ranging from 2.9% (1024x1024 array) to 10.4% (256x256 array) over previous CiM, considering full parallelism. However, if the peripherals can be shared amongst the columns (as in some architectures), the area overhead ranges from 0.34% to 3.16% (considering 4:1 column multiplexing). As an alternate design of the compute module, the two additional multiplexers can be replaced by an XOR and an AOI21 gate to obtain *both* subtraction and addition outputs. This design has a 4 transistor overhead compared to the former but enables CiM of addition and subtraction in the same cycle.

The *SUM* values of all the compute modules constitute the addition or subtraction outputs. The input carry bit $C_{IN}$ of the first stage is '0' for addition and '1' for subtraction. The *CARRY* of the previous stage is propagated as the $C_{IN}$ of the next stage. $n+1$ compute modules are used for every $n$ bit subtraction. The additional $(n+1)^{th}$ compute module is used for dealing with overflows and gets its $C_{IN}$ from the carry output *CARRY* of the $n^{th}$ compute module. Since the inputs being subtracted are in two's complement form and can be sign-extended, the other two inputs to the $(n+1)^{th}$ compute module are the same as the inputs to the $n^{th}$ compute module.

With in-memory subtraction, CiM of comparison can be achieved seamlessly. Since both the inputs and the output are in two's complement form, the value of the most significant bit of the subtraction output (SUM of the $(n+1)^{th}$ compute module) can reveal the larger of the two numbers. If both the inputs are equal, every bit of the output will be a zero, which can be detected by a near-memory AND gate tree. For an $n$ bit comparison, $n-1$ two-input AND gates are needed for the AND tree, adding an overhead of just 1 gate per column. Thus, the CiM of comparison is enabled through the CiM of subtraction with minimal overheads.

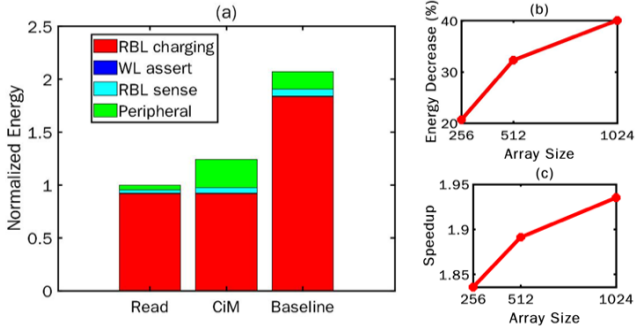

Figure 4: (a) energy components of the read, ADRA based CiM and the baseline, (b) energy decrease and (c) speedup of ADRA based CiM as a function of the array size for current based sensing

## IV. RESULTS

In this section, we analyze in-memory subtraction/comparison using ADRA in the context of both current and voltage-based sensing (since previous CiM works have considered both sensing schemes [3,4]) We evaluate the technique on a 1T FeFET NVM array using the following bias conditions: $V_{READ}$ (RBL) = 1V; $V_{GREAD} = V_{GREAD2}$ = 1V; $V_{GREAD1}$ = 0.83V; $V_{SET}$ = 3.7V and $V_{RESET}$ = -5V. With these values we obtained a sense margin of > 50mV and > 1µA for voltage and current-based sensing, respectively for ADRA. We evaluate speedup/energy benefits and trade-offs compared to the baseline, which needs two memory accesses and near-array subtraction (recall previous CiM approaches cannot perform a single cycle subtraction [8][13]).

### A. Current Based Sensing

Current based sensing directly converts the current produced on the senseline during read and CiM to the corresponding digital value with a current sense amplifier. Fig. 4(b) and 4(c) show the energy decrease and the speedup of ADRA-based CiM over the near-memory baseline as a function of the array size. At an array size of 1024 x 1024, the proposed CiM operation is 1.94x faster and uses 41.18% lesser energy than the baseline, resulting in an energy-delay product (EDP) decrease of 69.04%. The near 2x speedup is because ADRA based CiM and the near-memory baseline require 1 and 2 cycles for subtraction respectively. The energy benefits can be understood by examining the various energy components of the standard read and the proposed CiM operation. Fig. 4(a) shows the energy components per column in a 1024 x 1024 1T FeFET array for a 32-bit word corresponding to RBL charging/discharging, flow and sensing of read current, word line charging/discharging, and peripheral circuitry (sense amplifiers and compute module). The RBL charging is the primary energy component for both read (91% of total energy) and CiM (74% of total energy). Since the peripheral circuit energy is not a major component, the energy overheads of the additional hardware in the compute module leads to only a small increase in the overall energy. Our analysis shows that the CiM operation expends 1.24 times the energy of the standard read operation. However, subtraction in the baseline designs (i.e., without ADRA) would require two reads and a compute. Thus, ADRA based computation leads to a 41.18% decrease in energy. It is seen that both the speedup and the energy benefits of ADRA increase as the array size increases. This is because RBL charging/discharging energy becomes more dominant with an increase in array size, amortizing the energy and latency overheads of the ADRA peripherals.

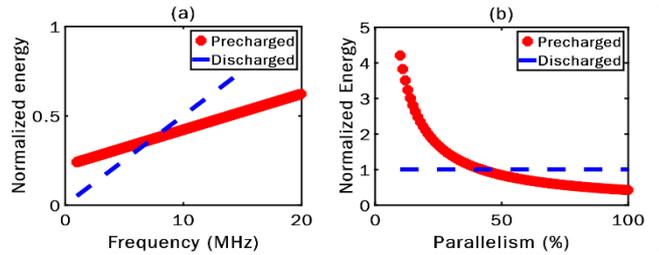

Figure 5: energy comparison for CiM using precharged (scheme 1) and discharged RBL (scheme 2) voltage sensing schemes as a (a) function of frequency of CiM operation and (b) parallelism of CiM operation

### B. Voltage Based Sensing

CiM architectures such as [4] and [13] have used voltage-based sensing for read and compute. Two versions of voltage-based sensing are commonly used: (1) keeping the read bitline (RBL) precharged to $V_{DD}$ during the hold operation and (2) discharging RBL to 0 during hold and charging it to $V_{DD}$ for every read/compute operation. We will refer to these as scheme 1 and scheme 2 respectively. In scheme 1, the RBL charging energy per operation is relatively small compared to scheme 2. However, the array expends leakage energy during the hold state due to the pre-charged RBLs. Therefore, for low frequency systems, where the number of CiM operations per second is low, the extra energy spent in scheme 2 for charging the bitlines would be balanced by the savings in the leakage energy. Fig. 5(a) displays this trade-off for ADRA showing that at frequencies below 7.53 MHz, scheme 2 is more energy efficient.

Another important aspect to consider when analyzing the two voltage sensing schemes is the parallelism in the CiM operation. Let us define the amount of parallelism $P$ as $N_{W,CiM}/N_{W,TOT}$ where $N_{W,CiM}$ is the number of words in a row on which the CiM operation is being performed parallelly and $N_{W,TOT}$ is the total number of words stored in a row. For example, CiM on a *single* word (used in some general purpose architectures) and on *all* words would correspond to $P=1/N_{W,TOT}$ and P = 1, respectively. Now, since the wordline is common to all the words in the row, the words not involved in the computation are still accessed. This is especially critical for scheme 1, in which the half-selected words effectively go through a 'pseudo CiM' operation (like pseudo-read [19]). Hence, some energy is wasted in charging the bitlines of the *half-selected* words, as they may be discharged during pseudo-CiM. In contrast, scheme 2 does not suffer from this energy overhead since only the *selected* RBLs are charged during read/compute. Thus, scheme 2 is more suitable for arrays with a low $P$ (<~42% in Fig. 5(b)).

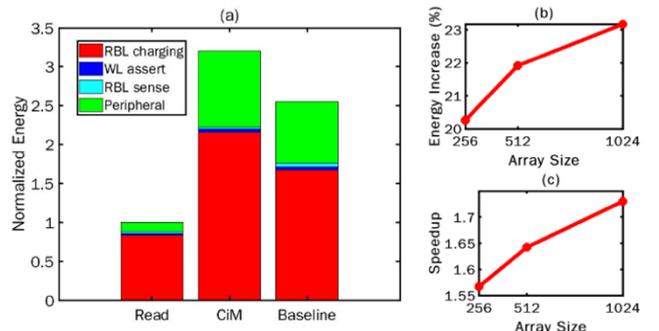

Figure 6: (a) energy components of read, ADRA based CiM and baseline, (b) energy decrease and (c) speedup of ADRA based CiM as a function of array size for precharged RBL voltage sensing (scheme 1)

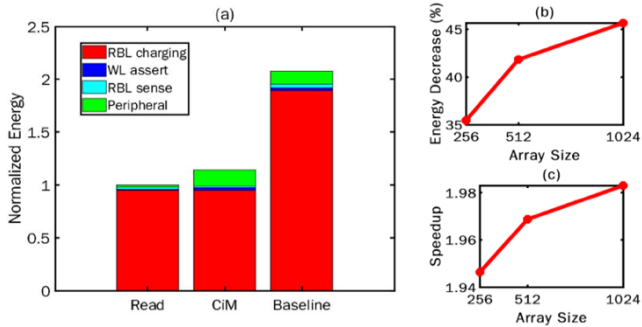

**Figure 7: (a) energy components of read, ADRA based CiM and baseline, (b) energy decrease and (c) speedup of ADRA based CiM as a function of array size for discharged RBL voltage sensing (scheme 2)**

Fig. 6(b) and 6(c) show the performance comparison of ADRA compared to the baseline for scheme 1. It is seen that the speedup ranges from 1.57x to 1.73x and increases with an increase in the array size, a trend similar to that seen in current based sensing. However, the CiM computation costs 20-23% more energy than the baseline. Examination of the energy components (Fig. 6(a)) of the standard read and the CiM operations shows that the bitline charging energy is the most dominant energy component for both read and CiM. In current based sensing, the bitline charging energy components for standard read and CiM have negligible difference. However, for scheme 1, the bitline charging energy for CiM is approximately 3 times that of the standard read. This can be explained as follows: Let Δ be the sense margin of the voltage sense amplifier. For the standard read operation (in a single ended memory), the bitline will have to discharge by 2Δ to reliably distinguish between the *two* possible values (0 and 1). However, in ADRA, the bitline will need to discharge by 6Δ to reliably distinguish between the *four* possible input vectors. Thus, compared to the baseline (two reads + compute), ADRA exhibits an energy overhead of 1.5X in the bitline energy component. However, since the peripheral circuit energies for both read and compute are similar, ADRA based CiM computation using scheme 1 has a net 20-23% energy overhead and 1.57x-1.73x speedup over the baseline, leading to a 23.26% - 28.81% decrease in EDP.

On the other hand, for scheme 2, the implications of ADRA are different (See Fig. 7 (a-c)). CiM using ADRA has a speedup of 94.5 - 98.3% and expends 35.5 - 45.8% lesser energy than the baseline, leading to a 66.83% - 72.6% decrease in EDP. This is because the RBLs are charged before the read/CiM, and hence, RBL energy component is similar for standard read and ADRA. Therefore, the effect of ADRA on the energy components and the overall performance and energy are similar to current sensing.

## V. CONCLUSION

In this paper, we propose ADRA, an asymmetric dual row assertion scheme which achieves a single-cycle 2 bit read, enabling full scale digital CiM of functions with two in-memory operands. ADRA maps each input vector to a unique current or voltage value on the senseline, allowing the CiM of non-commutative functions such as subtraction and comparison in a single cycle, in addition to the previously implemented commutative functions. We evaluate our technique on a 1T FeFET array and present its implications for a current-based and two voltage-based sensing schemes. We show that ADRA exhibits 23.2% - 72.6% decrease in EDP compared to the near-memory compute baseline.